\documentclass{rspublic}
\usepackage{graphicx}
\usepackage{amsmath}
\usepackage{amsfonts}
\usepackage{amssymb}
\usepackage{amssymb}
\usepackage{psfrag}
\usepackage{natbib}
\usepackage{color}

\newcommand{\beq}{\begin{equation}}
\newcommand{\eeq}{\end{equation}}
\newcommand{\Keff}{K_{\mathrm{eff}}}
\renewcommand{\r}{\mathbf{r}}
\newcommand{\Bext}{\mathbf{B}_{\mathrm{ext}}}
\newcommand{\m}{\mathbf{m}}
\newcommand{\upd}{\mathrm{d}}

\newcommand{\dl}{\Delta \ell}
\newcommand{\vinf}{v_\infty}
\newcommand{\tgamma}{\tilde{\gamma}}
\newcommand{\tf}{\tilde{f}}

\title{The \emph{magneto-elastica}: From self-buckling to self-assembly}

\author[D. Vella, E. du Pontavice, C. L. Hall \& A. Goriely]{Dominic Vella, Emmanuel du Pontavice, Cameron L.~Hall and Alain Goriely}

\affiliation{Mathematical Institute, University of Oxford, Oxford, OX2 6GG, UK}

\begin{document}

\maketitle

\begin{abstract}{elasticity, discrete-to-continuum asymptotics, self-assembly}
Spherical neodymium-iron-boron magnets are  perman-ent magnets that can be assembled into a variety of structures due to their high magnetic strength. A one-dimensional chain of these magnets  responds to mechanical loadings in a manner reminiscent of an elastic rod. We investigate the macroscopic mechanical properties of assemblies of ferromagnetic spheres by considering chains, rings,  and chiral cylinders of magnets. Based on energy estimates and simple experiments, we introduce an effective magnetic bending stiffness for a chain of magnets and show that, used in conjunction with classic results for elastic rods, it provides excellent estimates for the buckling and vibration dynamics of magnetic chains. We then use this estimate to understand the dynamic self-assembly of a cylinder from an initially straight chain of magnets.
\end{abstract}

\section{Introduction}

The self-assembly of magnetic particles is of great general scientific and engineering interest appearing as it does in a range of key applications from microscopic swimming \cite[][]{kuaral10}, through water filtration \cite[][]{yavuz06} to the manufacture of high-end electronic devices \cite[][]{sumuwe00}, and the design of new synthetic viruses \cite[][]{pesisa03}. The general scientific problem is to understand  the process by which magnetic particles self-assemble through a combination of magnetic interactions and other mechanical forces, and to explore the influence of the initial distribution of magnets and random fluctuations on the process of assembly. However, once formed, the overall macroscopic properties of the assembly are also of considerable interest. Previous research has  focused on modeling two specific cases: chains of paramagnetic beads linked by a molecular bond \cite[][]{dreyfus05,roper06} or the preferred arrangement of large ensembles of dipolar nano-particles using both experiment and computation  \cite[][]{ku10,vabeka13}.

%%%%%%%%%%%%%%% End of first page %%%%%%%%%%%%%%%%%%%%%

\noindent In this article, we consider the collective elastic behavior of interacting dipolar ferromagnetic spheres arranged in a chain, and we study the spontaneous self-assembly of such a chain into a chiral cylinder.

\begin{figure}
\centering
\includegraphics[width=0.9\columnwidth]{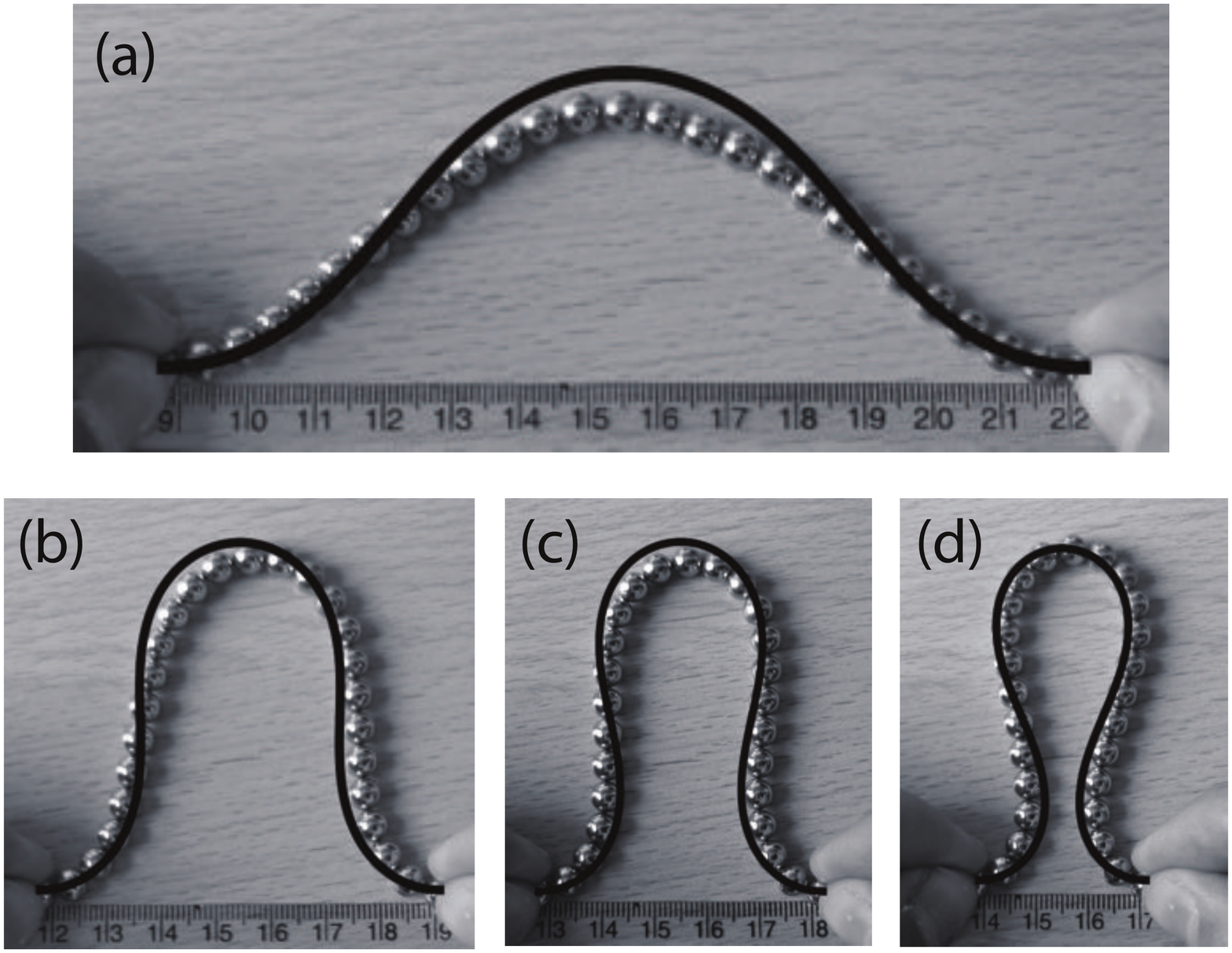}
\caption{Confinement of the \emph{magneto-elastica}. A chain consisting of $N=25$ spheres (with diameter $2a=6\mathrm{~mm}$ and magnetic field strength $B=1.195\mathrm{~T}$) is compressed by bringing its ends closer by a distance $\Delta L$ and forms a shape that is compared to the elastica with the same end-end compression (solid curves). Results are shown for (a) $\Delta L/2aN=0.19$, (b) $\Delta L/2aN=0.56$, (c) $\Delta L/2aN=0.68$ and (d) $\Delta L/2aN=0.80$.}
\label{ElasticaShapes}
\end{figure}

We use collections of  millimetre-sized spherical neodynium-iron-boron (NdFeB) magnets as a paradigm for dipolar self-assembly. These magnets can  be easily obtained either as toys (under the brand names `Neocube' or `Buckyballs') or as well-calibrated magnets for engineering applications. The high magnetic strength of these spheres can be used to create various shapes and structures that resist  mechanical deformation through their magnetic interaction. The simplest demonstration of this resistance to deformation is seen by taking a chain of beads and bringing the ends closer together (see figure \ref{ElasticaShapes}). The deformed chain retains a coherent shape, which is remarkably reminiscent of the classic elastica that is formed by performing the same experiment with an elastic rod or beam \cite[][]{love}. Because of this similarity we refer to this shape as the `\emph{magneto-elastica}', though we stress that the spheres do not have any elastic connection in the usual sense. 

\begin{figure}
\centering
\includegraphics[width=0.9\columnwidth]{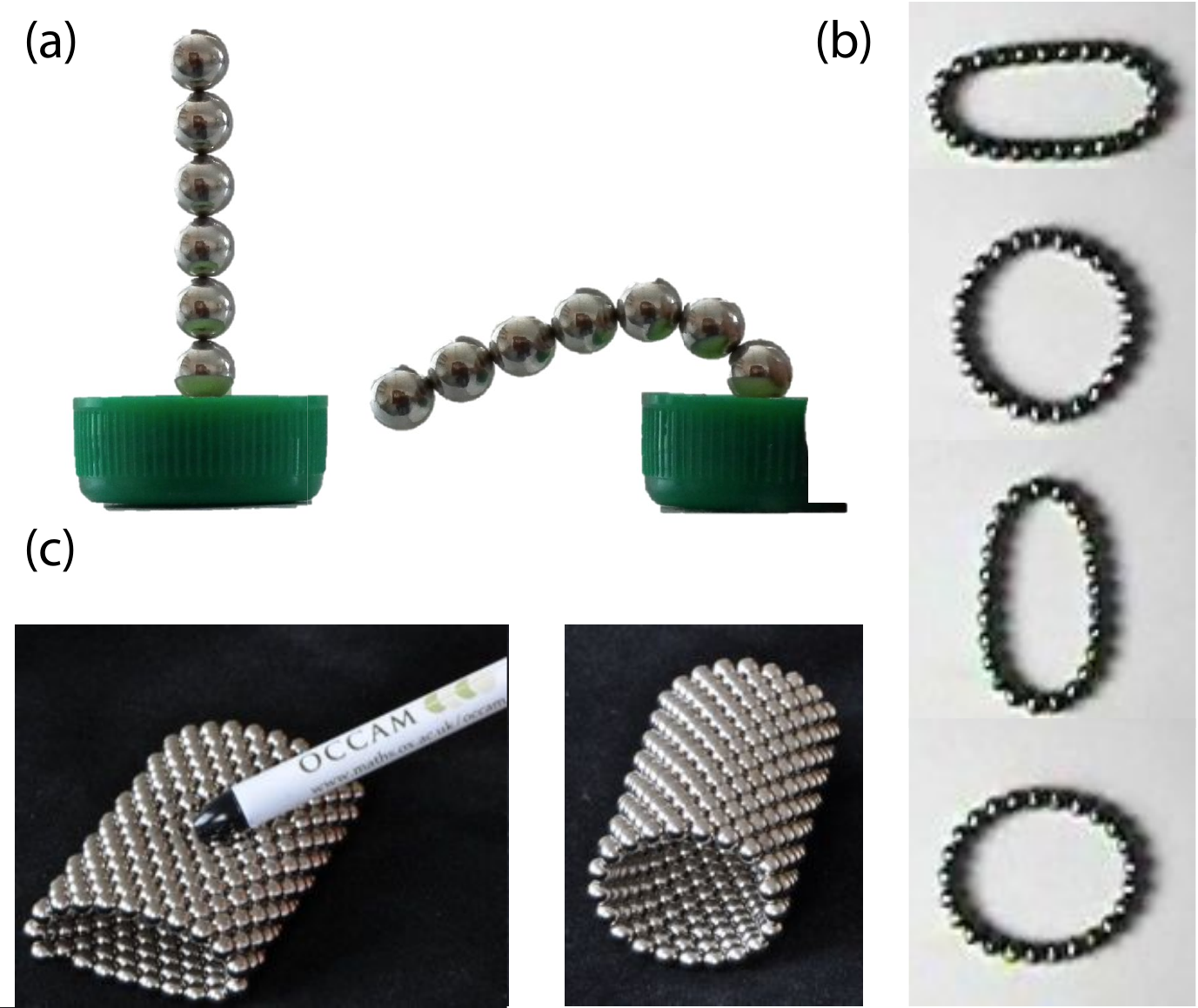}
\caption{Three simple experiments that illustrate the resistance to deformation of assemblies of ferromagnetic spheres. (a) The self-buckling of a vertical chain of magnetic spheres as further spheres are added. (b) The prolate-oblate oscillation of a ring of magnetic spheres. Snap-shots of the motion are shown at intervals of $0.021\mathrm{~s}$. (c) A self-assembled chiral `nano-tube'. In each case, the diameter of the spheres is $2a=5\mathrm{~mm}$ and the magnetic field strength is $B=1.195 \mathrm{~T}$.}\label{Fig1}
\end{figure}

Instead, the spheres will interact through their magnetic fields, with the consequence that magnetic beads that are far apart along the chain but close to each other in space will potentially have strong interactions that cannot be accounted for using classical elastica theory. These interactions between widely-spaced spheres (along with other issues such as the discreteness of the chain and imperfections in the application of `clamped' boundary conditions)  may explain some of the discrepancy observed in Figure \ref{ElasticaShapes} between the confinement of the magneto-elastica and the results of classical elastic theory. A detailed analysis of the differences between the magneto-elastica and the classical elastica would involve the difficulty of quantifying the differences between two shapes. However, the qualitative similarity seen in figure \ref{ElasticaShapes} is striking, suggesting that the concept of an effective `bending stiffness' for the magneto-elastica may be a useful one. Further demonstrations of the elastic-like nature of the magneto-elastica can readily be found. For example, chains held vertically buckle under their own weight, a closed ring oscillates, and a cylinder resists bending but recovers its shape after poking (figure~\ref{Fig1}). As described below, these experiments give us opportunities to test quantitatively the hypothesis that the magneto-elastica can be approximated by a chain with a magnetism-induced bending stiffness.

\begin{figure}
\centering
\includegraphics[width=0.9\columnwidth]{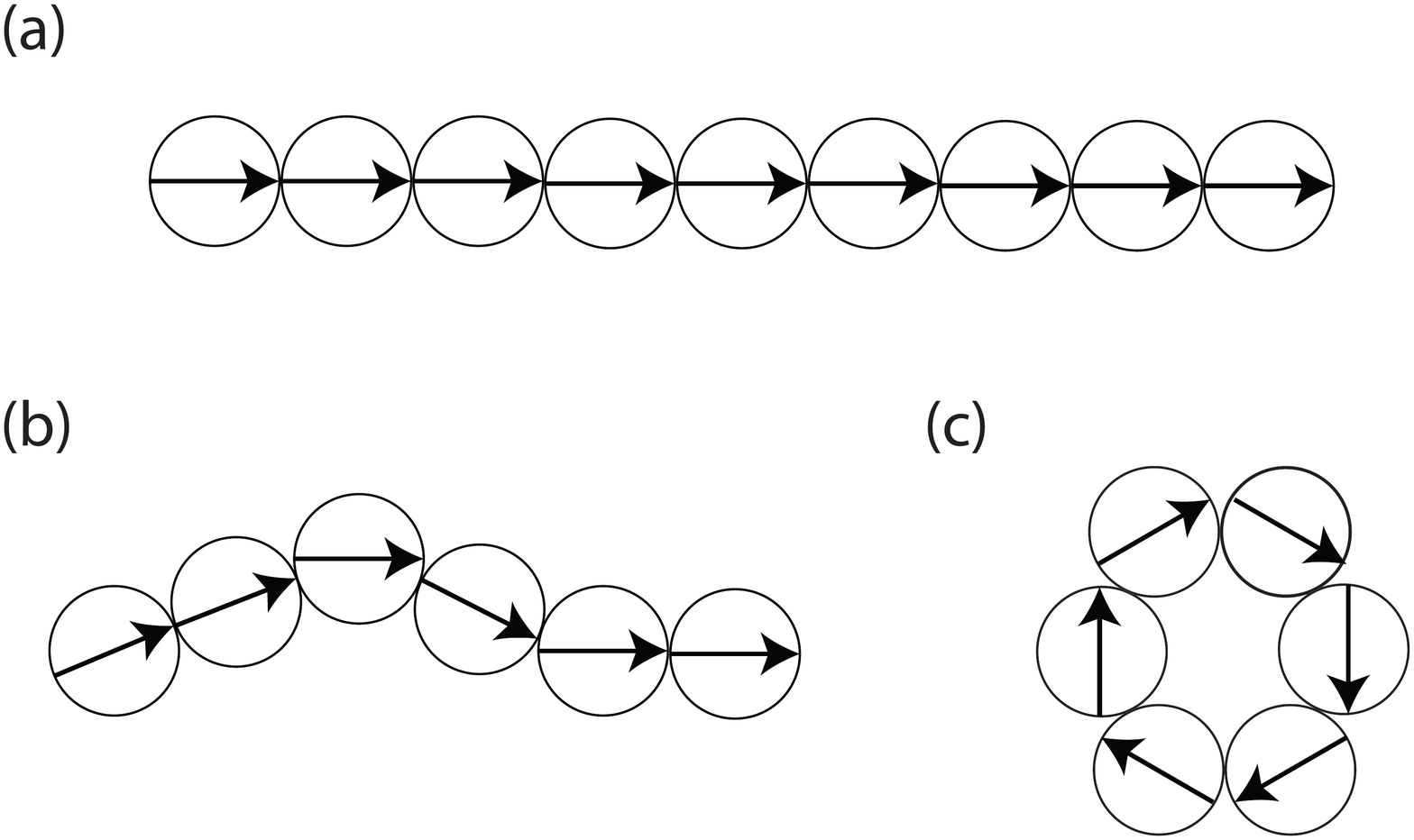}
\caption{The physical mechanism behind the resistance to deformation of a magnetic chain. (a) In a straight chain all of the magnetic dipoles are aligned. (b) When deformed the dipoles can no longer be aligned. This frustration gives rise to a torque that resists deformation. (c) The general problem of determining the dipole orientation for a given deformation is in principle complicated \cite[][]{hall13}; here we consider a model calculation based on computing the energy required to form  closed rings from chains.}
\label{fig:dipoles}
\end{figure}

The origin of the resistance to deformation exhibited in figure \ref{Fig1} is simple to understand in physical terms. For simplicity, we assume that the magnetic spheres have a uniform internal magnetisation and recall that  the external magnetic field around such a sphere is precisely dipolar \cite[][]{jackson}. At equilibrium, a straight chain of spherical magnets is therefore an oriented collection of magnetic dipoles (shown schematically in figure \ref{fig:dipoles}a). However, when this chain is forced to bend, the dipoles cannot be aligned both with the chain and with each other (see figure \ref{fig:dipoles}b). In this frustrated configuration, the magnetic field of each dipole exerts a torque on every other dipole. The combined resistive torque that is generated acts to  straighten the chain locally and reduce its curvature. This behaviour reminds us of an elastic rod, which exerts a torque proportional to the curvature $\kappa$ to resist bending. However, this torque resisting bending  is different from 
the `tension' that has previously been observed in a chain of paramagnetic spheres placed in an external magnetic field \cite[][]{dreyfus05,roper06}.

For an elastic rod, the ratio of the bending moment applied and the curvature induced, $\kappa$, is called the bending stiffness, which is frequently denoted $K$. Provided that the radius of curvature is large compared to the typical cross-sectional dimension of the rod, $K$ is a constant, and, for an elastic rod of radius $a$ and Young's modulus $E$, it is well-known that
\beq
K=\frac{\pi}{4}Ea^4.
\label{eqn:rodbendstiffness}
\eeq The bending stiffness has the dimensions of a force times a length squared, i.e.~$\mathrm{Nm^2}$. Based on such dimensional considerations alone we might expect that a magnetic chain should have an effective bending stiffness of the form
\beq
\Keff=\frac{B^2a^4}{\mu_0} f(a\kappa,N).
\label{eqn:keff}
\eeq Here $B$ is the magnetic field strength (a quantity that is specified by the manufacturer in Tesla where $1\mathrm{~T}=1\mathrm{~N A^{-1}m^{-1}}$), $\mu_0=4\pi\times10^{-7}\mathrm{~NA^{-2}}$ is the permeability of free space, $a$ is the radius of an individual sphere and $f(a\kappa,N)$ is a dimensionless function of the dimensionless chain curvature, $a\kappa$, and the number of spheres in the chain, $N$.

It is not clear \emph{a priori} whether the function $f(a\kappa,N)$ in \eqref{eqn:keff} is a constant, how it might depend on $N$ or even why $\Keff$ should not depend on derivatives of the curvature.  An analysis of this question for general shapes has been considered in detail elsewhere \cite[][]{hall13} and shows that there is, in fact, a non-local contribution to the deformation energy that cannot be explained as a result of an ``effective bending stiffness". However, this analysis is complicated to the extent that even deriving the equilibrium equations for the shape of a chain in all but some very simple cases is impossible. In this paper, therefore, we adopt a more direct approach: we compute the energy required to deform a linear magnetic chain into a polygonal ring and, by comparing to the corresponding elastic result determine the corresponding dimensionless bending stiffness. In this simplified case symmetry indicates that the function $f(a\kappa,N)$ of \eqref{eqn:keff} is a function of $N$ alone, $\tf(N)$. Moreover, we find that $\tf(N)$ rapidly approaches a constant value as $N$ increases. Having shown this we then proceed to use the concept of a magnetic bending stiffness to understand quantitatively some different experimental scenarios, including those shown in figure \ref{Fig1}a,b.

\section{The effective bending stiffness\label{sec:effbend}}

\subsection{Theoretical background}

We begin by considering a single magnetic sphere centred at the origin and recall that, assuming the internal magnetisation is uniform, then the external field  at a position $\mathbf{r}$ due to this sphere is \cite[][]{jackson}
\beq
\mathbf{B}(\r)=\frac{\mu_0}{4\pi}\frac{3(\m\cdot\r)\r-r^2\m}{||\mathbf{r} ||^5}.
\eeq Here the dipole moment of a sphere of radius $a$ is $\m=\frac{4\pi a^3}{3}M\hat{\m}$ with $M$ the strength of the internal (uniform) magnetization and $\hat{\m}$ a unit vector in the direction of the dipole moment. (Note that for the magnetic spheres available commercially the value of $B=\mu_0M$ is given by the manufacturers and is typically in the range  1.1 to 1.4 Tesla.)
%In the models developed for chains of paramagnetic spheres,  it has been shown that the chain is subject to an induced tension resisting deformation of the chain from straight.
%For chains of ferromagnetic spheres it quickly becomes clear that there is also a resistance to bending. Intuitively, it seems clear that the torque that resists bending should depend on the curvature of the chain (since it is the tilting of dipole moments with respect to one another that matters, rather than their global orientation). 
The approach we adopt here is based on the calculation of  the magnetic energy of various configurations of spheres. The energy of a single magnetic dipole $\m$ in an externally imposed magnetic field, $\Bext$ is \cite[][]{jackson}
\beq
U=-\m\cdot\Bext.
\eeq 
Since the external field of a magnetic sphere is precisely the same as a dipole with dipole moment $\m$, the energy of interaction between two magnetic spheres must be precisely the energy of two dipoles with the same strength, separation and orientation \cite[][]{hall13}. Therefore, the total energy of a collection of $N$ magnetic spheres located at $\mathbf{r}_i$ with dipole moment $\mathbf{m}_i$ is
\begin{equation}
\label{UN}
  U_{N} 
 = -\sum_{i=1}^N \, \sum_{\substack{j=1 \\ j\neq i}}^{N} \frac{\mathbf{m}_i \cdot \mathbf{B}_j(\mathbf{r}_i)}{2},
\end{equation}
where
\[
 \mathbf{B}_j(\mathbf{r}) = \frac{\mu_0}{4 \, \pi} \, \frac{3 \, \big[(\mathbf{r} - \mathbf{r}_{_j}) \cdot \mathbf{m}_j\big] \, (\mathbf{r} - \mathbf{r}_j) - ||\mathbf{r} - \mathbf{r}_j||^2 \, \mathbf{m}_j}{||\mathbf{r} - \mathbf{r}_j||^5}
\] and the factor $1/2$ is introduced in \eqref{UN} to ensure that the interaction energy is not double-counted.

We shall proceed by considering the energies of a chain consisting of $N$ spheres in a linear chain and in a closed ring\footnote{These configurations are chosen because their symmetry dictates that the dipole moments must be aligned with the tangent vector of the deformed chain. In fact, this result holds more generally for chains whose curvature is not too large, but involves a detailed calculation that is not particularly enlightening \cite[][]{hall13}.}. Having calculated these energies we evaluate their asymptotic behaviour in the limit of large chains, $N\gg1$, by making extensive use of the Euler--Maclaurin formula \cite[][]{knopp90,olver10}, which are given for completeness in Appendix \ref{sec:eulermac}.

\subsection{A linear chain}

Before calculating the energy of a finite linear chain, see figure \ref{fig:dipoles}a, we note that it is already known, in fact, that the closed ring is energetically favourable in comparison to the linear chain for all $N>3$ \cite[][]{prokopieva09,vandewalle13}. This is because of the release of energy that occurs when the two spheres at the end of such a chain are brought close to one another; as such this difference in energy does not reflect the energy required to frustrate the dipole alignment, which is the basis of the stiffness that is of interest to us here. Rather than considering the energy of a finite linear chain, therefore, we instead consider the energy of a line of $N$ spheres embedded within an infinite chain. Orientating the infinite chain along the $y$-axis, the position vector of the centre of the $i-$th bead is
\beq
\r_i=2a(0,i), \quad i\in \mathbb{Z}.
\eeq Clearly, in equilibrium the dipole moments must be parallel to the line of the chain; without loss of generality we assume that the moments are orientated in the negative direction so that the dipole moments of the beads are given by
\beq
\m_i=(0,-1)\frac{4\pi a^3}{3}M, \quad i\in \mathbb{Z}.
\eeq

We consider the energy of the bead at the origin in the magnetic field of all of the other beads, $U_0$. The energy of a chain of length $N$ will thus be $U_0\times N/2$, with the factor of $2$ included to avoid double counting the energy, and the simple form to go from a single sphere to all $N$ arising from the fact that the outer chain is infinite.

Now, the magnetic field at the origin is
\beq
\mathbf{B}(\r=0)=\frac{\mu_0}{4\pi}\sum_{{i=-\infty\atop i\neq0}}^\infty\frac{3(\r_i\cdot\m_i)\r_i-r_i^2\m_i}{r_i^5}.
\eeq This will only have a component in the $y$ direction (and indeed we are only interested in this component since $\mathbf{m}_0\propto \mathbf{j}$); we therefore calculate
\beq
\mathbf{j}\cdot \mathbf{B}(\r=0)=\frac{\mu_0}{4\pi}\sum_{{i=-\infty\atop i\neq0}}^\infty\frac{-16\pi a^5 Mi^2+4a^2i^2\frac{4\pi a^3}{3}M}{(2a|i|)^5}=-\frac{\zeta(3)}{6}\mu_0M,
\eeq where $\zeta(3)=\sum_{i=1}^\infty i^{-3}\approx 1.202$. Hence
\beq
U_0=-\m_0\cdot\mathbf{B}(\r=0)=-\frac{2\pi\zeta(3)}{9}\mu_0a^3M^2
\eeq and the energy of the chain of $N$ such magnets is
\beq
U_N^{(\mathrm{chain})}=\frac{N}{2}U_0=-\frac{\pi\zeta(3) N}{9}\mu_0a^3M^2.
\label{eqn:energyLine}
\eeq

\subsection{A finite ring}

To isolate the effects of bending, we consider a polygonal ring of $N$ spheres, with the centre of the polygon being at the origin. As $N\to\infty$ this approaches a circle, and so will isolate any curvature-dependent energy that arises.

Letting $R=a\csc(\pi/N)$ be the distance of the centre of each magnetic sphere from the origin, we have that the position vector of these centres is
\beq
\r_i=R\left[\cos\frac{2\pi}{N}(i-1),\sin\frac{2\pi}{N}(i-1)\right],\quad i=1,2,..., N
\eeq and, again taking the bead $i=1$ to be orientated in the negative $y$ sense, the magnetic moments are
\beq
\m_i=\frac{4\pi a^3}{3}M\left[\sin\frac{2\pi}{N}(i-1),-\cos\frac{2\pi}{N}(i-1)\right],\quad i=1,2,..., N
\label{eqn:miring}
\eeq since symmetry dictates that $\m_i\cdot\r_i=0$.

Using symmetry again, it is enough to calculate the energy of the bead $i=1$ and multiply this value by $N/2$ (again to avoid double counting). To do this, we note that
\beq
|\r_1-\r_i|=2R\sin\frac{\pi}{N}(i-1).
\eeq  We find that
\beq
U_1=-\frac{\pi}{18}\frac{a^6M^2}{R^3}\sum_{i=2}^N\frac{1+\cos^2\frac{\pi}{N}(i-1)}{\sin^3\frac{\pi}{N}(i-1)}
\eeq  and hence
\beq
\frac{U_N^{(\mathrm{ring})}}{\mu_0M^2a^3}=-\frac{\pi}{36}N\sin^3\frac{\pi}{N}\sum_{j=1}^{N-1}\frac{1+\cos^2\frac{\pi}{N}j}{\sin^3\frac{\pi}{N}j}.
\label{eqn:energyring}
\eeq We note that an equivalent expression has been found previously in studies of the ground state of dipolar particles \cite[][]{prokopieva09}.

 In deriving the expression \eqref{eqn:energyring}, we have not made any approximations. However, to progress further we need to evaluate the sum
\beq
S=\sum_{j=1}^{N-1}\frac{1+\cos^2\frac{\pi}{N}j}{\sin^3\frac{\pi}{N}j}=2\sum_{j=1}^{(N-1)/2}\frac{1+\cos^2\frac{\pi}{N}j}{\sin^3\frac{\pi}{N}j}+O(1)=2H+O(1)
\eeq in the limit $N\gg1$. To do this, we  would like to apply the Euler--Maclaurin formula \eqref{EulerMaclaurin}. However, the summand in $H$ (and hence also its derivatives) grows without bound as $jN^{-1} \rightarrow 0$. In order to overcome this difficulty, we notice that, for $j$ close to $1$, the summand in $H$ may be approximated by a Taylor series for large $N$ (or small $\pi/N$). We therefore write $H=H_1+H_2$ where
\beq
H_1=\sum_{j=1}^{N^\alpha-1}\frac{1+\cos^2\frac{\pi}{N}j}{\sin^3\frac{\pi}{N}j},\quad
H_2=\sum_{j=N^\alpha}^{(N-1)/2}\frac{1+\cos^2\frac{\pi}{N}j}{\sin^3\frac{\pi}{N}j}
\eeq and $0<\alpha<1$.

Now, expanding the summand in $H_1$ for $\pi j/N\ll1$ (which holds for $N\gg1$ since $j\leq N^\alpha<N$), we have
\beq
H_1=\sum_{j=1}^{N^\alpha-1}\left(\frac{2N^3}{\pi^3j^3}+\frac{7\pi}{60 N}j+...\right)=\frac{2N^3}{\pi^3}\left(\zeta(3)-\sum_{j=N^\alpha}^\infty j^{-3}\right)+\frac{7\pi}{120}N^{2\alpha-1}+O(1).
\eeq The partial sum of $j^{-3}$ in this expression may be computed using the Euler--Maclaurin formula, which gives
\beq
H_1=\frac{2\zeta(3)}{\pi^3}N^3-\frac{N^{3-2\alpha}}{\pi^3}-\frac{N^{3(1-\alpha)}}{\pi^3}+\frac{7\pi}{120}N^{2\alpha-1}+O(1).
\eeq

To evaluate $H_2$, we immediately make use of the Euler--Maclaurin formula, which yields
\beq
H_2= \frac{N}{\pi}\int_{\pi N^{\alpha-1}}^{\pi/2-\pi/2N}\frac{1+\cos^2y}{\sin^3y}~\upd y+\frac{1}{\pi^3}N^{3(1-\alpha)}+O(1),
\eeq where the constant terms occur from the evaluation of the function at the limits of the integral. Calculating the integral analytically and performing Taylor series expansions about $y=0$ and $y=\pi/2$, we find that
\beq
H_2= \frac{N}{\pi}\left[-\frac{\pi}{2N}+O(N^{-3})+\frac{N^{2(1-\alpha)}}{\pi^2}-\frac{1}{6}-\frac{7}{120}\pi^2N^{2(\alpha-1)}\right]+\frac{1}{\pi^3}N^{3(1-\alpha)}+O(1).
\eeq Combining this result with that for $H_1$, we see that the terms involving $\alpha$  all cancel (as they must, since $\alpha$ was arbitrary) and we are left with
\beq
S=\frac{4\zeta(3)}{\pi^3}N^3-\frac{1}{3\pi}N+O(1).
\eeq

Using this result in \eqref{eqn:energyring} we find that
\beq
\frac{U_N^{(\mathrm{ring})}}{\mu_0M^2a^3}=-\frac{\pi}{36}N\frac{\pi^3}{N^3}\left(1-\frac{\pi^2}{2N^2}+...\right)\times \frac{4\zeta(3)N^3}{\pi^3}\left(1-\frac{\pi^2}{12\zeta(3)N^2}\right)+O(1),
\eeq or
\beq
\frac{U_N^{(\mathrm{ring})}}{\mu_0M^2a^3}=-\frac{\pi\zeta(3)}{9}N+\frac{\pi^3}{18}\left[\zeta(3)+\frac{1}{6}\right]N^{-1}+O(N^{-2}).
\label{eqn:energyringasy}
\eeq

\subsection{Energy due to bending and the effective bending stiffness}

In determining the energy of deformation, it is the difference in energies between the deformed and undeformed states that is of interest. Comparing the ring and the linear chain, we find that the ring has an energy that is higher by an amount
\beq
\frac{\Delta U_N}{\mu_0M^2a^3}=\frac{U_N^{(\mathrm{ring})}-U_N^{(\mathrm{chain})}}{\mu_0M^2a^3}=\frac{\pi^3}{18}\left[\zeta(3)+\frac{1}{6}\right]N^{-1}+O(N^{-2}).
\label{eqn:energyDeformation}
\eeq 

As expected, the ring state is energetically unfavourable compared to the straight configuration --- there is an energy penalty to be paid by bending the chain into a closed loop. The energy change can be computed exactly for different values of $N$ by evaluating the sum in \eqref{eqn:energyring} numerically, which gives the energy of a closed ring, and comparing it to the energy of a linear chain of the same length, which is given by \eqref{eqn:energyLine}. This numerical computation of the exact energy difference between the deformed and undeformed configurations shows that the closed loop is energetically unfavourable for all $N\geq 4$, but that for $N=3$, the closed loop, i.e.~triangular, configuration is actually  energetically favourable compared to the straight chain.

\begin{figure}
\centering
\includegraphics[width=0.49\columnwidth]{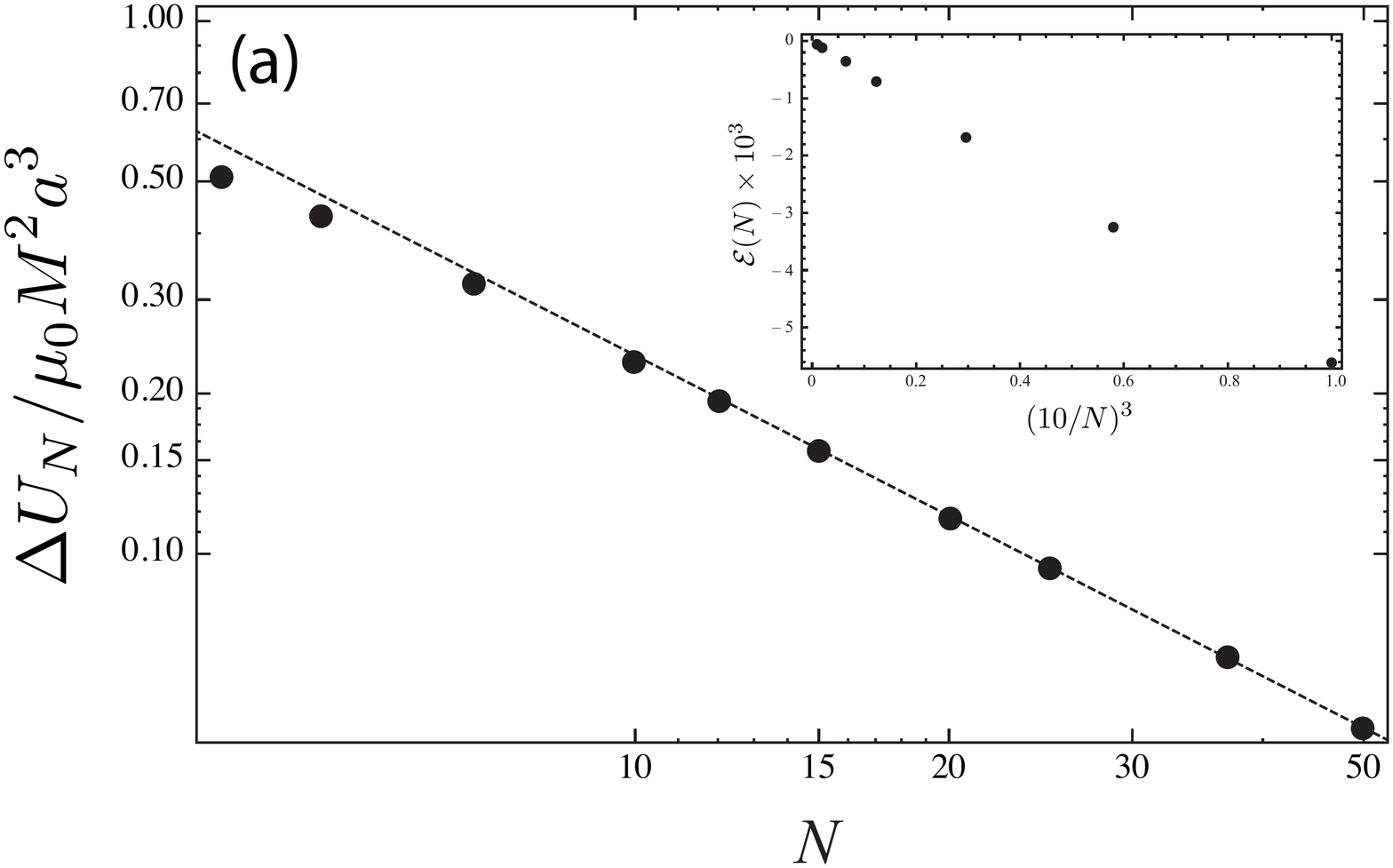}
\includegraphics[width=0.49\columnwidth]{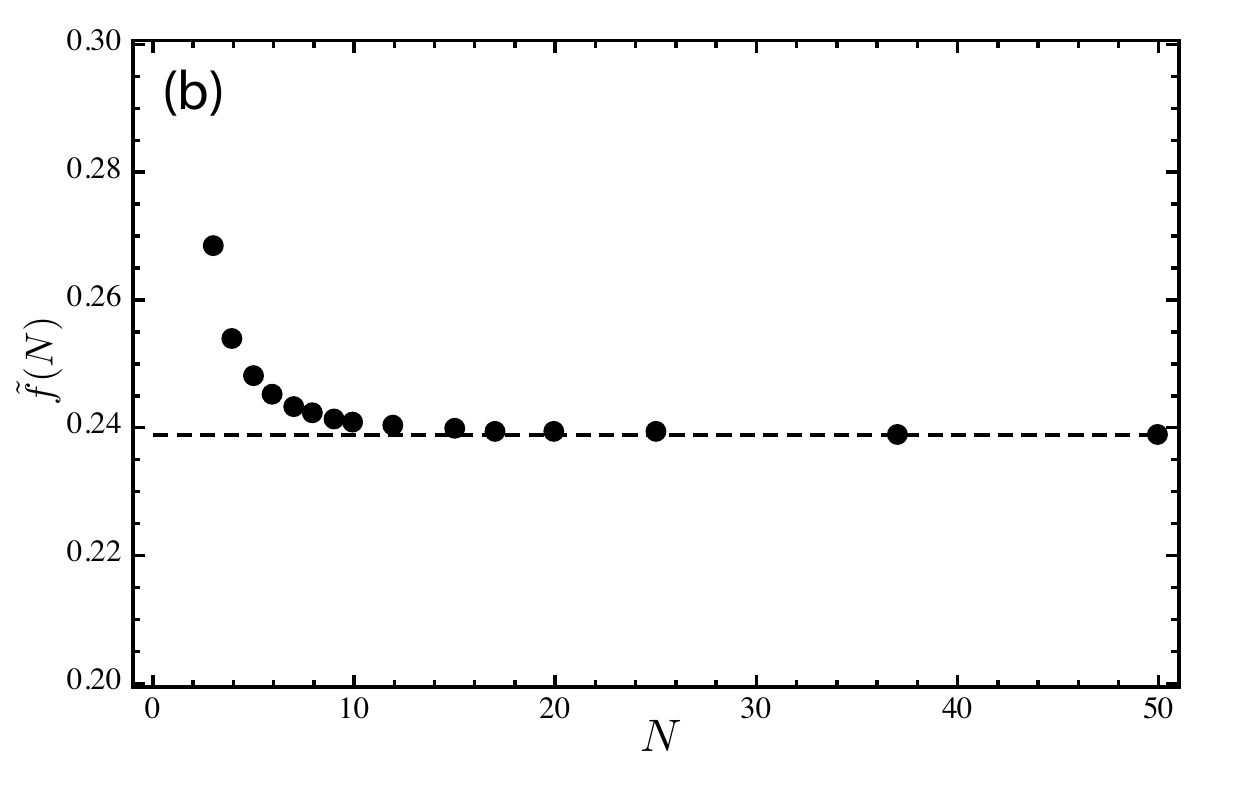}
\caption{Numerical and asymptotic results for the effective bending stiffness. (a) The numerically computed change in energy $\Delta U_N$ as a function of $N$ (points) rapidly tends to the asymptotic result \eqref{eqn:energyDeformation} (dashed line). Although  the error estimate in \eqref{eqn:energyDeformation} leads us to expect that ${\cal E}(N)\sim N^{-2}$, numerical computations reveals that in fact ${\cal E}(N)\sim N^{-3}$ (see inset). (b) The behaviour of the dimensionless bending stiffness $\tf(N)=f(\sin\tfrac{\pi}{N},N)$, with $f$ as defined in \eqref{eqn:keff}, computed numerically (points) and in the limit $N\gg1$ (dashed line).}
\label{fig:deltaUN}
\end{figure}

Figure \ref{fig:deltaUN}a shows the asymptotic behaviour of $\Delta U_N$ for large $N$. We observe that the leading order behaviour of the asymptotic result \eqref{eqn:energyDeformation} is in excellent agreement with numerical computations of the change in energy for even quite moderate $N\gtrsim10$. Furthermore, the error in \eqref{eqn:energyDeformation} appears to be $O(N^{-3})$, although we are only able to formally justify that this error should be $O(N^{-2})$. This suggests that a cancellation occurs in the calculation of the $N^{-2}$ term in \eqref{eqn:energyDeformation} that we are not able to access using the relatively simple calculation outlined here.

Finally, we compare the energy of deformation in \eqref{eqn:energyDeformation} with that of an elastic rod of bending stiffness $K$ and length $L$ that is bent round to form a circular ring of radius $R=L/2\pi$. The energy of an elastic ring is simply $U_B=K/2\times L/R^2=2\pi^2 K/L$. By identifying this energy with the energy \eqref{eqn:energyDeformation} of our magnetic ring of length $L= 2aN$,  we obtain an effective bending stiffness
\beq
K_\mathrm{eff}=\mu_0M^2a^4 \frac{\pi}{18}\left[\zeta(3)+\frac{1}{6}\right]\approx0.239\frac{B^2a^4}{\mu_0}.
\label{eqn:bendstiffness}
\eeq Compared to an elastic rod  with a circular cross section, radius $a$ and bending stiffness $K=E\pi a^{4}/4$  the effective bending stiffness in \eqref{eqn:bendstiffness} corresponds to an effective Young's Modulus $E_\mathrm{eff}\approx0.956 B^2/\mu_0\pi\approx 400\mathrm{~kPa}$; our magnetic solid is a rather soft material.

We note that the form of \eqref{eqn:bendstiffness} is identical to that given on purely dimensional grounds in \eqref{eqn:keff}. However, we also note that the dimensional analysis that led to \eqref{eqn:keff} was not able to tie down the dimensionless function $f$ or indeed to determine whether it should depend on $N$. In fact, we have found that, to leading order in $N^{-1}$, the bending stiffness is a constant independent of $N$. This result is compared to numerical results in figure \ref{fig:deltaUN}b where we see that, as expected, the numerically computed $\tf(N)\sim  \tfrac{\pi}{18}\left[\zeta(3)+1/6\right]$ as $N\to\infty$.  While this notion of an effective bending stiffness only holds exactly for the closed rings considered here\cite[][]{hall13}, the remainder of this paper is concerned with determining whether this concept is useful in gaining quantitative understanding of some more general scenarios beyond that for which the above analysis is strictly valid.

\section{The heavy \emph{magneto-elastica}}

A  vertical column of magnetic spheres will buckle once its weight reaches a critical value, as shown in figure \ref{Fig1}a. A purely numerical approach to this problem is to compute the shape that minimizes the gravitational and magnetic energies of a finite chain of spheres; such an approach reveals that sufficiently short chains remain vertical, while the chain buckles above a critical length (i.e.~number of spheres), as is also observed experimentally. However, this minimization process does not highlight the underlying physics at play in such scenarios. A complementary approach is to use the classic result \cite[][]{wang86} that a heavy elastic rod with bending stiffness $K$ and linear mass density $\rho_\ell$ buckles when its length 
\beq
L\geq L_c\approx 1.986(K/g \rho_\ell)^{1/3}.
\eeq For the chains of magnetic spheres considered here, $L=2aN$ and $\rho_\ell=2\pi  \rho_s a^2 /3$. Using the effective bending stiffness $K_\mathrm{eff}$ from \eqref{eqn:bendstiffness} we thus expect that buckling will occur if
\beq
N>N_c\approx 0.4817{\cal G}^{-1/3}
\label{eqn:critNbuckling}
\eeq where we have introduced the Magneto--Gravitational number
\beq
{\cal G}=\frac{\mu_0\rho ga}{B^2}
\eeq to characterize the relative importance of the magnetic properties and the weight of the spheres.

\begin{figure}
\centering
%\psfrag{g}{${\cal G}$}
%\psfrag{n}{$N$}
\includegraphics[width=0.90\columnwidth]{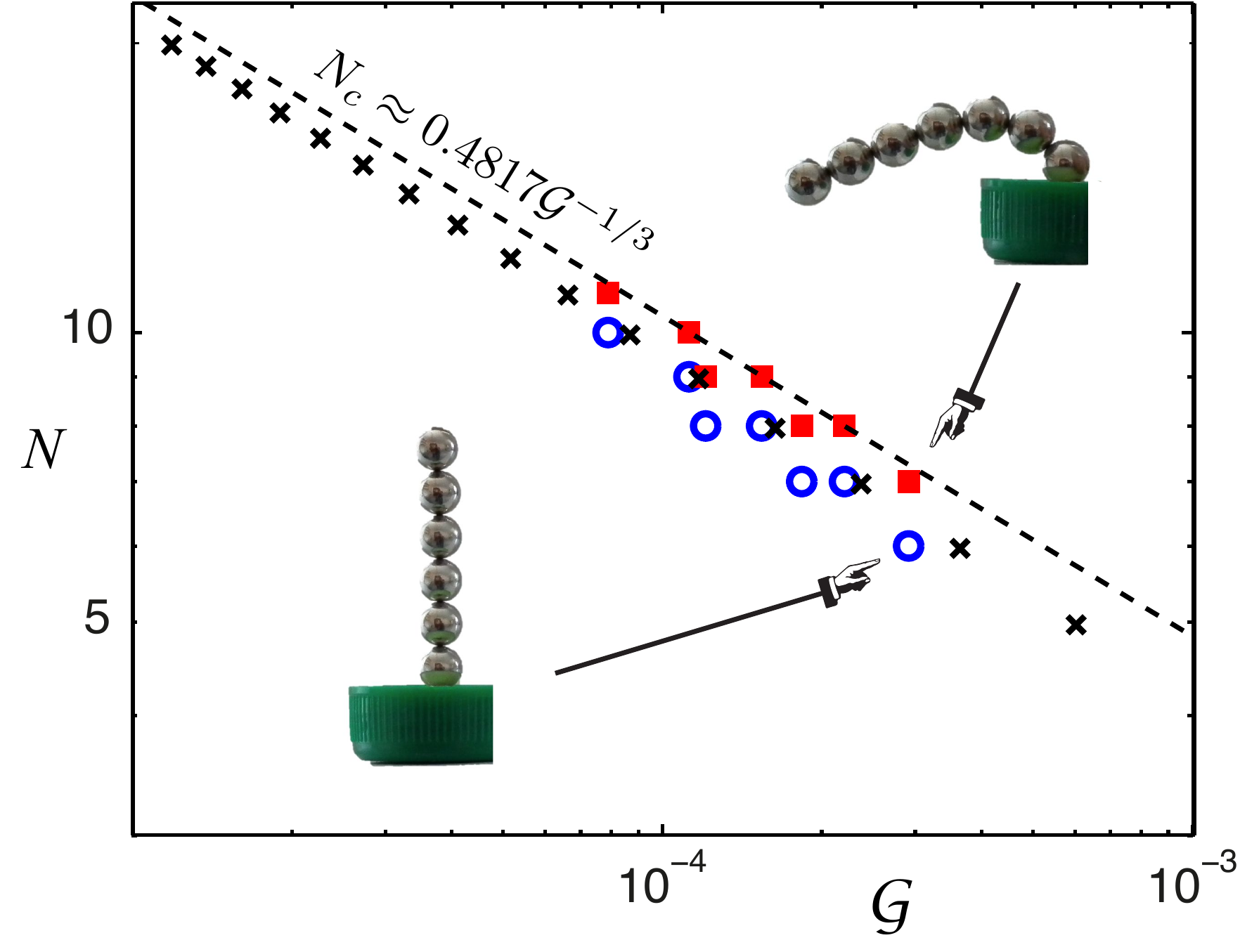}
\caption{Experimental, numerical and asymptotic results showing the regions of $({\cal G},N)$ parameter space for the buckling of a chain of magnets under its own weight. Squares correspond to experiments in which the chain buckled while circles correspond to experiments in which the chain remained straight. Crosses correspond to the results of direct numerical computations  (using the optimization toolbox of MATLAB) which, for a given number of spheres, gives the critical Magneto--Elastic number ${\cal G}_c$ at which buckling occurs. The dashed line gives the corresponding approximate relationship \eqref{eqn:critNbuckling}, which is obtained using the idea of an effective magnetic bending stiffness and the classic result for the buckling of a heavy elastic column \cite[][]{wang86}.}
\label{fig:critNbuckling}
\end{figure}

Experiments with spheres of different radii $a$ in the range $1.5\mathrm{~mm}\leq a\leq 4\mathrm{~mm}$  and different magnetic strengths $1.195\mathrm{~T}\leq B \leq 1.4\mathrm{~T}$ were conducted to determine the number of spheres required for self-buckling (with a vertical clamp at the base). (In these, and the other experiments described in this paper, we do not measure the value of $B$ directly but rather assume that the value given by the manufacturer is correct.) In these experiments, a chain that is sufficiently short to stand vertically despite gravity is first formed. Additional spheres are then added to the top of the chain and whether the chain remains straight or buckles is noted. The largest (respectively smallest) values of $N$ for which the chain remained vertical (respectively buckled) are presented for a number of values of the Magneto--Gravitational number ${\cal G}$ in Fig.~\ref{fig:critNbuckling}. We observe that the scaling for the critical number of spheres at which buckling occurs, \eqref{eqn:critNbuckling}, is in good agreement with experiments while the pre-factor given in \eqref{eqn:critNbuckling} overestimates $N_c$ by around $10\%$.

To investigate whether the difference between experiments and theory observed in fig.~\ref{fig:critNbuckling} is due to experimental uncertainties or rather the errors that are inevitably introduced in applying a result derived for a continuous, elastic problem to this discrete scenario, we also performed numerical computations of the discrete problem. In these computations, $N$ identical spheres are aligned in a linear configuration that is slightly tilted with respect to the vertical (at an angle $\pi/100$) and with their magnetic dipoles aligned in the same direction. The optimisation toolbox in MATLAB was then used to find the minimum energy configuration of these spheres by adjusting their position  and the orientation of their magnetic moments. For a given number of spheres, this leads to a numerically computed value of the critical Magneto--Gravitational number, ${\cal G}_c$, at which buckling occurs. These results are quantitatively consistent with experiments, which were performed with a fixed ${\cal G}$ (Fig.~\ref{fig:critNbuckling}). This suggests that the error between the approximate result \eqref{eqn:critNbuckling} and experiment is due to the approximations made in deriving this result (e.g.~the discreteness of the experimental system and long range interactions between spheres), rather than inaccuracies in experiments. Nevertheless, the concept of a bending stiffness that is purely magnetic in origin gives a reasonable quantitative understanding of experiments.

\section{Two dynamic scenarios}

Having seen the reasonable success of the theory of the classic \emph{elastica} to determine the behaviour of a chain of magnetic spheres in two static scenarios, it is natural to ask whether the idea of an effective bending stiffness is also useful in understanding some dynamic scenarios. This is the subject of this section.

\subsection{Oscillating rings}

A classic dynamic demonstration of the restoring force due to elasticity is to pinch an elastic ring and observe the ensuing oscillations. This classic problem was first considered by Hoppe \cite[][]{love,hoppe71} who showed that, for an elastic ring with bending stiffness $K$, radius $R$ and linear density $\rho_\ell$, the oscillation frequency of the  $n$-th mode is given by
\beq
\omega_n^2=\frac{EI}{\rho_\ell R^4} \frac{n^2(n^2-1)^2}{n^2+1}.
\eeq

Performing the same experiment with a magnetic chain is simple (see figure \ref{Fig1}b for some snap shots of this motion); a chain with $N$ spheres in it will form a ring of radius $R\approx Na/\pi$ and will have $\rho_\ell=2\pi\rho a^2/3$. The experiment considered here corresponds to the prolate-oblate oscillation mode, i.e.~$n=2$, and so, after substituting $K_{\mathrm{eff}}$ from \eqref{eqn:bendstiffness}, we expect that
\beq
\omega=\pi^2\left\{ \frac{3}{5}\left[\zeta(3)+\frac{1}{6}\right]\right\}^{1/2}\frac{B}{(\rho\mu_0)^{1/2} a N^2},
\label{eqn:oscringfrequencyLove}
\eeq or, alternatively, in dimensionless form
\beq
\Omega\equiv\frac{\left(\mu_0\rho\right)^{1/2}a\omega}{B}=\pi^2\left\{\frac{3}{5}\left[\zeta(3)+\frac{1}{6}\right]\right\}^{1/2}N^{-2}.
\label{eqn:oscringNDfreqLove}
\eeq

\begin{figure}
\centering
\includegraphics[width=0.90\columnwidth]{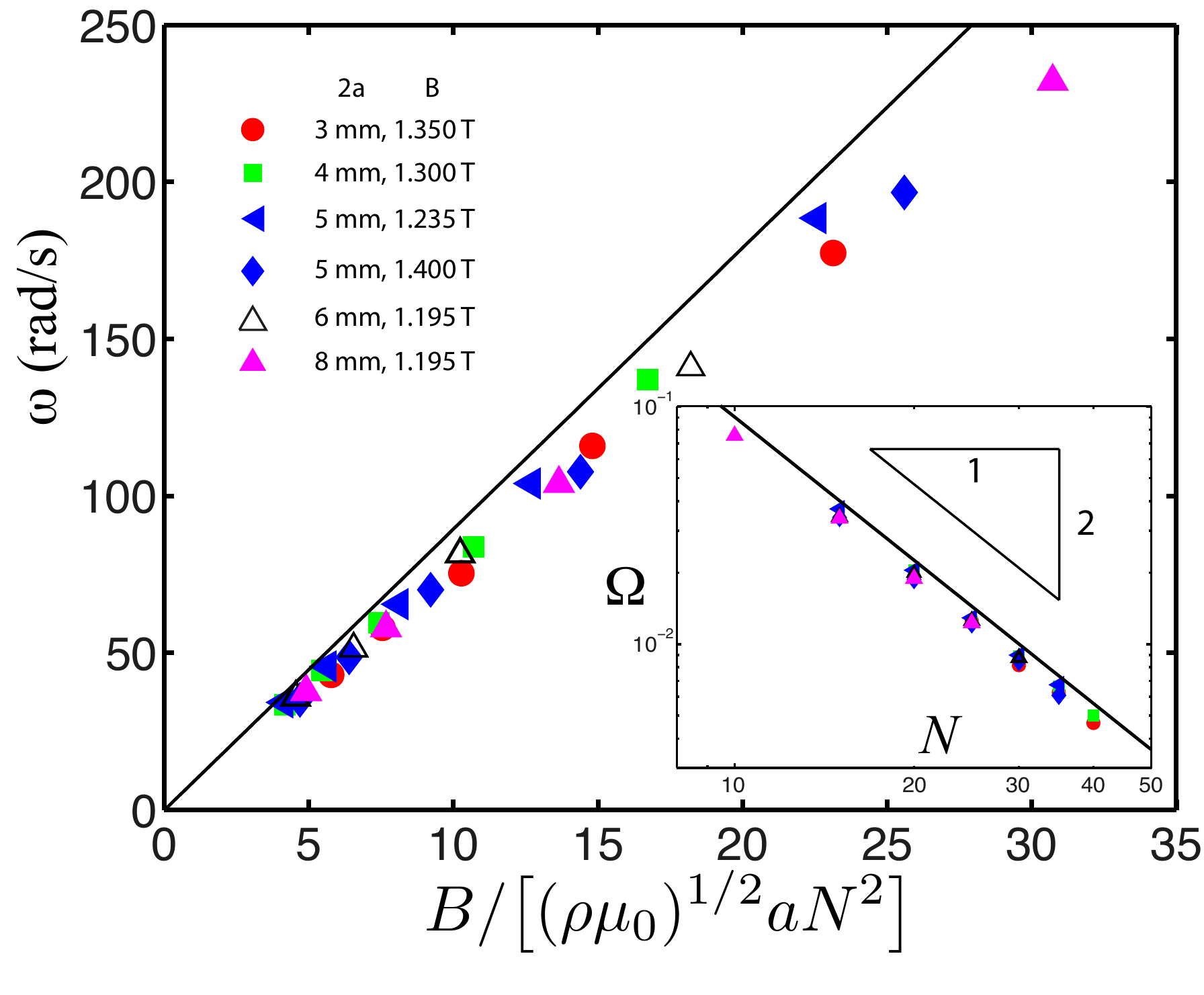}
\caption{Experimental and theoretical results for the natural frequency of oscillation of a ring of magnetic spheres. Main figure: The relationship between the frequency $\omega$ and the material parameters observed experimentally (points) is as predicted by \eqref{eqn:oscringfrequencyLove} (solid line). Inset: The dimensionless oscillation frequency $\Omega$ predicted by \eqref{eqn:oscringNDfreqLove}, solid line, is also borne out by experiments (points). The symbols used to encode different sphere sizes and magnetizations are as described in the legend.}
\label{fig:OscFreqResults}
\end{figure}

To test these results, we performed a series of experiments in which magnetic chains of different lengths, sphere radii and magnetic strengths are formed into rings and placed on a smooth horizontal table. The rings are then pinched and released. The frequency $\omega$ of the resulting oscillations was measured from high speed video footage obtained using a Finepix HS10 camera at frame rates of up to $240\mathrm{~Hz}$. The experimental results are plotted in figure \ref{fig:OscFreqResults}. We see excellent quantitative agreement between experiments and the simple theoretical analysis, which is based on the idea of an effective magnetic bending stiffness developed in \S\ref{sec:effbend}.

\subsection{A self-assembling cylinder}

\begin{figure}
\centering
%\psfrag{x}[c]{$\frac{B}{(\mu_0\rho)^{1/2}aN^2}$}
%\psfrag{n}{$N$}
%\psfrag{w}[c]{$\omega \mathrm{ ~ (rad/s)}$}
%\psfrag{v}{$\Omega$}
\includegraphics[width=0.90\columnwidth]{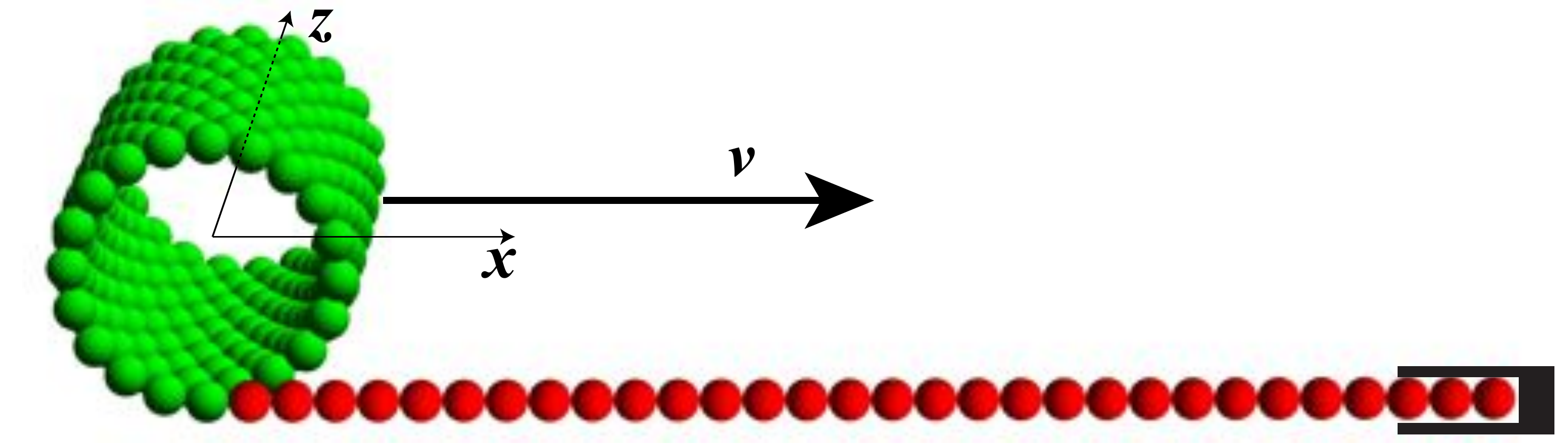}
\caption{Self-assembly of a cylindrical helix from a chain. The straight chain is clamped on the right side and exerts both a torque and a force on the cylinder. For sufficiently large `seeding' cylinders, the cylinder spontaneously rolls up the chain increasing the length (but not the radius) of the cylinder.}
\label{fig:cylinder-exp}
\end{figure}

As a final example of the utility of the notion of an effective magnetic bending stiffness  developed in this article we consider a problem that has, to our knowledge, no classic analogue in the theory of elasticity. This experiment is shown schematically in figure \ref{fig:cylinder-exp}: a long chain of $M+P$ spheres is laid on a horizontal surface with one end clamped and the other end rolled into a small cylindrical helix. The cylindrical portion, containing $P$ spheres, has radius $R(N)$, and there are $N$ spheres in each circuit of the cylinder  (so that any sphere $k$ is in contact with its neighbours $k\pm1$ and the spheres $k\pm N$ and $k\pm(N+1)$ along the helix). The straight chain initially consists of $M$ magnets lying on a table along the $x$-axis (see Fig.~\ref{fig:cylinder-exp}). For `seeding' cylinders of sufficiently large diameter, this configuration is unstable, and the straight portion of the chain spontaneously wraps itself onto the cylinder extending  the initial cylindrical helix (see 
Supplementary Information for movies of this process).

\subsubsection{A simple scaling law\label{sec:cylscale}}

A simple scaling argument, using the idea of a bending stiffness developed in \S\ref{sec:effbend} allows us to gain some understanding of this self-assembly phenomenon. We begin by noting that self-assembly occurs because it is energetically favourable for the spheres within the chain to be aligned with, and brought closer to, their neighbours; this is achieved by wrapping up into a cylinder. However, to do this they must go from being in a linear configuration (with dipoles aligned) to being in a curved configuration (with dipoles partially frustrated). Any excess energy gain that remains once this penalty has been paid can increase the kinetic energy of the sphere. While in the initial stages of the motion this energy gain may also be used to accelerate the whole cylinder, at late times we expect the cylinder to be travelling at a steady speed, $\vinf$, so that the change in energy is converted solely into kinetic energy.

Consider therefore an element of the linear chain $\dl$ as it is assimilated into the helix. We assume that the energy released as the element is taken from the chain and brought into contact with its neighbours is some constant, $\gamma$, per unit length; the total energy released for this element is then just $\gamma\dl$. However, there is a bending energy penalty, $\tfrac{1}{2}\Keff R(N)^{-2}\dl$, and any remaining energy will be used to give this element a kinetic energy $\rho_\ell\vinf^2\dl$ (assuming the cylinder rolls, its kinetic energy is $M\vinf^2$, rather than $\tfrac{1}{2}M\vinf^2$). Equating these energies we therefore expect that
\beq
\rho_\ell\vinf^2\approx\gamma-\frac{\pi^2}{2}\frac{\Keff}{N^2 a^2},
\eeq which may be written
\beq
\vinf^2\sim \frac{B^2}{\mu_0\rho}\left( \tgamma-\frac{1}{N^2}\right)=\frac{B^2}{\mu_0\rho}\left(\frac{1}{N_c^2} -\frac{1}{N^2}\right).
\label{eqn:v2scale}
\eeq In \eqref{eqn:v2scale}, $\tgamma=1/N_c^2$ represents a dimensionless adhesive energy gain and the significance of $N_c$ will become apparent shortly.

There are two interesting, surprising and important features of the scaling law \eqref{eqn:v2scale} that we discuss now, before going on to repeat a more detailed version of this calculation. Firstly, we notice that the speed appears to depend only on the number of spheres in each winding of the seeding cylinder, $N$, and the strength of the magnetic field, $B$; in particular, $\vinf$ is independent of the sphere radius, $a$. Secondly, $\vinf$ becomes imaginary if $N<N_c$ so that $N_c$ may be interpreted as the most tightly wound cylinder that can still self-assemble. Though surprising on first reading, this latter result makes physical sense since, if the cylinder is too tightly wound, the energy penalty to be paid upon winding up into a helix is prohibitive in comparison to the energy that is released by coming into contact with one's neighbours.

 \subsubsection{A more detailed calculation}

We now estimate the forces acting on the spheres in the helical configuration. If the straight portion of the chain is long enough ($M>15$) and the number of helical repeats is large enough ($P/N>6$), the force $F_{x}$ and torque $\tau_{z}$ acting on the cylinder from the chain are well-approximated by constants. We assume that the cylinder rolls without slipping and that its velocity is purely along the $x$-axis (as defined in figure \ref{fig:cylinder-exp}). Since the cylinder increases in mass as it rolls,  it will reach a terminal velocity given by $\vinf^{2}=(F_{x}+\tau_{z}/R)/2 \rho_{l}$. Equivalently, from an energy conservation principle, the energy before and after a sphere is transferred from the line  to the cylinder is
\begin{equation}
U_M^{\mathrm{straight}}+U_{N,P}^{\mathrm{cyl}}+E_{P}^{\mathrm{cyl}}
=U_{M-1}^{\mathrm{straight}}+U_{N,P+1}^{\mathrm{cyl}}+E_{P+1}^{\mathrm{cyl}}
\end{equation}
where $E_{P}^{\mathrm{cyl}}=P \rho \tfrac{4}{3}\pi a^{3} \vinf^{2} $ is the kinetic energy of the rolling cylinder and $U_{N,P}^{\mathrm{cyl}}$ is the total magnetic energy of a chain of $P$ spheres wound into a helical cylinder  with $N$ spheres in each winding. This energy  can be estimated asymptotically for $N\gg1$ by realizing that the cylinder tends to a hexagonal lattice as $N$ tends to infinity and that, to order $N^{-1}$, the energy required to bend a straight chain is equivalent to the bending energy of the ring. The magnetic energy density (per sphere) in an infinite hexagonal lattice can be evaluated numerically to be 
$-\mu_0M^2a^3\pi\zeta(3)(2+\alpha)/18$ where $\alpha\approx 0.295$.  Therefore, we have
\begin{eqnarray}
&&\frac{U_{N,P}^{\mathrm{cyl}}}{\mu_0M^2a^3}=\\ \nonumber
&&-\frac{P\pi}{18 N}\left[\zeta(3)(2+\alpha)N-\pi^{2}(\zeta(3)+\tfrac{1}{6})N^{-1}+O(N^{-2})\right].
\end{eqnarray} (Note that the the $O(N^{-1})$ term here, which corresponds to the bending stiffness discussed earlier, is not affected by the change in geometry from a ring to a helix.)
Assuming a perfect motion along the $x$-axis without sliding and in the absence of other forms of friction, the steady velocity of the cylinder is 
\begin{equation}\label{predict}
\vinf^{2}\frac{\mu_0\rho}{B^2}=\tfrac{ 1}{144} \left[6 \alpha  \zeta (3)-\pi^2N^{-2} (6 \zeta (3)+1)+O(N^{-3})\right].
\end{equation} Note that the structure of \eqref{predict} is identical to that determined from the scaling analysis of \S4(b)\ref{sec:cylscale}, particularly \eqref{eqn:v2scale}. Of course, this detailed analysis allows us to determine a numerical value for $\tgamma=1/N_c^2$; we find that  $\tgamma\approx6.17^{-2}$ and so $N=7$ is the minimal number of spheres in each turn of the helix before it can roll, independent of all physical parameters. We again note that both \eqref{predict} and \eqref{eqn:v2scale}  predict a linear scaling of the velocity squared with $N^{-2}$. We now turn to some experimental tests of these predictions.

\subsubsection{Experimental results}

\begin{figure}
\centering
\includegraphics[width=0.95\columnwidth]{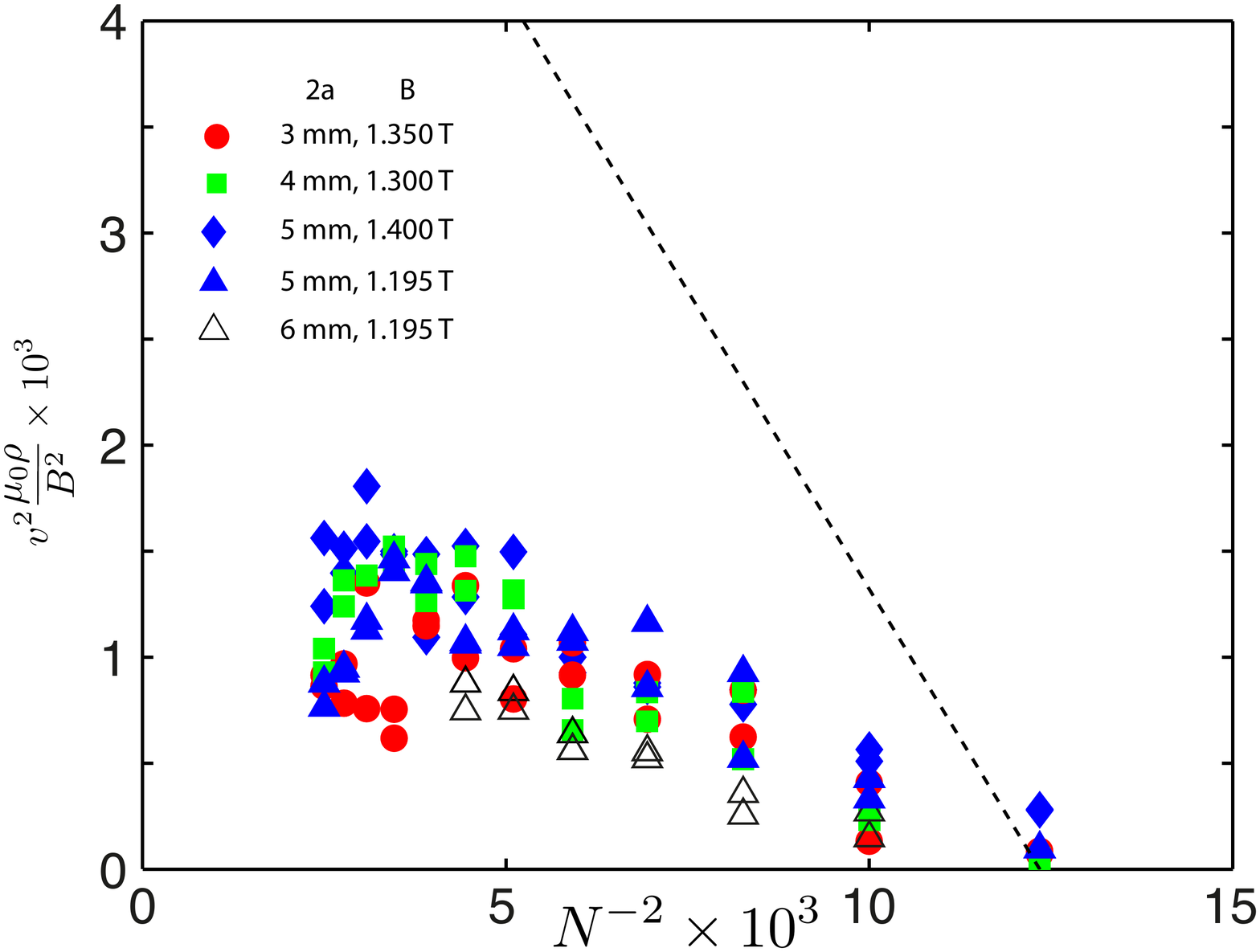}
\caption{Rescaled square of the terminal velocity of a self-assembling cylinder as a function of $N^{-2}$. As described in the text, the squared velocity scales linearly with  $N^{-2}$. The dotted line is the theoretical prediction \eqref{predict} with the pre factor $\alpha$ modified by the efficiency ${\cal E}$ (see main text). The symbols used to encode different sphere sizes and magnetizations are as described in the legend.}
\label{fig:cylinder}
\end{figure}

We performed experiments for helices made of spheres with a variety of sizes and  magnetic strengths as well as varying the size of the seeding cylinder. We do indeed observe that helices only self-assemble if the radius of the cylinder is sufficiently large, as predicted by both the scaling argument, \eqref{eqn:v2scale}, and the more detailed calculation that led to \eqref{predict}. However, contrary to the prediction of \eqref{predict} that self assembly will occur provided that $N\geq7$, we find experimentally that $N\geq9\pm1$ is required. In figure \ref{fig:cylinder} we observe a linear relationship between $v^2$ and $1/N^2$ and very little dependence on the size of the spheres, despite a doubling of the sphere diameter; both of these observations are in agreement with our theoretical arguments.

However, our theoretical estimate of the speed is about 3-4 times too large since it does not take into account the many dissipative effects present: the rolling of a cylinder made of discrete spheres (raising and lowering the centre of mass at each step), the rolling friction with the table and between the spheres, the audible noise generated as new spheres are added to the cylinder (see Supplementary Information for a video with soundtrack demonstrating this noise), the motion of the straight chain on the table along the $z$-axis, the force component along the $y$-axis, and so on. Rather than modelling the details of these dissipative effects, we use the experimentally observed number for rolling $N_{c}\approx9$ to estimate that the efficiency of conversion of magnetic energy into kinetic energy as spheres are brought into the helix is ${\cal E}\approx(6.17/9)^{2}=47\%$. Modifying \eqref{predict} by setting $\alpha\to\alpha{\cal E}$ we see that our theoretical estimate captures the main trends of the data, 
but still overestimates significantly, with $v_\text{theo}\approx2v_\text{expo}$. We attribute the remaining discrepancy to the fact that our estimate of dissipative effects accounts only for effects that prevent the cylinder from rolling up in the first place (cf.~static friction) while many dynamic dissipative effects also exist.

\section{Conclusions}

Systems of interacting spherical magnets constitute a conceptually simple and experimentally testable physical system for which all the interactions between the constituent units are known. The question addressed here, fundamental to many problems in physics, is to relate the macroscopic behaviour to the microscopic interactions. Indeed, Cauchy and Poisson related the macroscopic elasticity of crystals to the microscopic interactions between molecules \cite[][]{love}. Here, we have presented an example of this process by considering the simplest possible configurations of a chain of dipoles and determined a macroscopic effective bending stiffness. While the fact that the magnetic dipoles in a magnetic chain should resist bending of the chain is obvious (see figure \ref{fig:dipoles}), the analogy is only mathematically exact in certain very specialised geometries \cite[][]{hall13}. We have therefore investigated the utility of the simple physical notion of an effective `bending stiffness' by attempting to understand 
three simple experiments using it. We have found that the concept of an effective bending stiffness gives excellent results for the problems of the buckling of a standing column and the oscillations of a ring, while providing a qualitative understanding and the key features for the self-assembly of a chain into a cylindrical helix. However, detailed asymptotic calculations for other configurations \cite[][]{hall13} show that, in fact, this is not the whole story: interactions between widely spaced spheres can also play a role. 

The idea of a magnetic bending stiffness may have consequences for modelling the mechanics of chains of ferromagnetic particles along the lines of those models developed previously for chains of paramagnetic particles \cite[][]{dreyfus05,roper06}. However, we also highlight the possibility that the heavy \emph{magneto--elastica} and oscillating ring experiments may be useful in their own right as  simple assays through which the strength of spherical permanent magnets may be determined. In particular, the oscillation frequency of a circular ring grows linearly with the field strength $B$ and so is quite sensitive to variations in $B$. Similarly, the number of beads that can be formed into a straight vertical chain without buckling under its weight grows like $B^{2/3}$ and so is also sensitive to variations in $B$.

A natural generalization of the questions addressed here is the elasticity of two-dimensional sheets of magnets for which the same ideas can be applied. Moreover, the mathematical technique of approximating sums to obtain energies is applicable to a wide range of problems in other branches of materials science, from self-assembly of particles to understanding the mechanical properties of crystals. Not only do we expect that these mathematical approaches to particle interactions will have broad applications in the physical sciences, but we have confidence that, in particular, systems of magnets will become both a testing ground for new ideas in physics and a perfect demonstration tool for classical physical problems.

\section*{Acknowledgment}

This publication is based in part upon work supported by Award No. KUK-C1-013-04, made by King Abdullah University of Science and Technology (KAUST). AG is a Wolfson/Royal Society Merit Award Holder and acknowledges support from a Reintegration Grant under EC Framework VII. AG and CH acknowledges support   support from the EPSRC through Grant No. EP/I017070/1.

\appendix{The Euler--Maclaurin formula\label{sec:eulermac}}

 Given a function, $f \in \mathcal{C}^{2p}[m\epsilon,n\epsilon]$, the Euler--Maclaurin formula gives the following asymptotic result as $\epsilon \rightarrow 0$:
\begin{multline}
\sum_{i=m}^n f\left(i \epsilon\right)
=\epsilon^{-1}\int_{m \epsilon}^{n \epsilon} f(x)\mathrm{~d}x
-B_1\bigl[f(m \epsilon)+f(n \epsilon)\bigr] \\
+\sum_{k=1}^p\frac{B_{2k} \epsilon^{2k-1}}{(2k)!}\left[f^{(2k-1)}(n\epsilon)-f^{(2k-1)}(m\epsilon)\right]
+R \label{EulerMaclaurin}
\end{multline} where $B_i$ are the Bernoulli numbers, $B_1=-1/2$, $B_2=1/6, B_3=0,...$; $n - m$ is an increasing function of $\epsilon$; and $R$ is a remainder term given by
\begin{equation}
 R = \epsilon^{2p}\int_{m\epsilon}^{n\epsilon} P_{2p}\left( \left\{ \frac{x - m \epsilon}{\epsilon} \right\} \right) \, f^{(2p)}(x) \, \mathrm{d}x,
\end{equation}
where $P_i(x)$ are the Bernoulli polynomials, $P_1(x) = x - \tfrac{1}{2}$, $P_2(x) = x^2 - x + \tfrac{1}{6}$, \emph{etc.}~and $\{x\}=x-\lfloor x\rfloor$ is the fractional part of $x$.

%\bibliographystyle{jfm} 
%\bibliography{magnetchains} 

\begin{thebibliography}{17}
\expandafter\ifx\csname natexlab\endcsname\relax\def\natexlab#1{#1}\fi

\bibitem[Dreyfus {\em et~al.\/}(2005)Dreyfus, Baudry, Roper, Fermigier, Stone
  \& Bibette]{dreyfus05}
{\sc Dreyfus, R., Baudry, J., Roper, M.~L., Fermigier, M., Stone, H.~A. \&
  Bibette, J.} 2005 Microscopic artificial swimmers. {\em Nature\/} {\bf 436},
  862--865.

\bibitem[Hall {\em et~al.\/}(2013)Hall, Vella \& Goriely]{hall13}
{\sc Hall, C.~L., Vella, D. \& Goriely, A.} 2013 The mechanics of a chain or
  ring of spherical magnets. {\em {SIAM} J. Appl. Math.\/} {\bf 73}, 2029--2054.

\bibitem[Hoppe(1871)]{hoppe71}
{\sc Hoppe, R.} 1871 Vibrationen eines ringes in seiner ebene. {\em J. Reine
  Angewand. Math.\/} {\bf 73}, 158--170.

\bibitem[Jackson(1999)]{jackson}
{\sc Jackson, J.~D.} 1999 {\em Classical Electrodynamics\/}. Wiley.

\bibitem[Knopp(1990)]{knopp90}
{\sc Knopp, K.} 1990 {\em Theory and application of infinite series\/}. Dover.

\bibitem[Ku {\em et~al.\/}(2010{\natexlab{{\em a\/}}})Ku, Aruguete, Alivisatos
  \& Geissler]{kuaral10}
{\sc Ku, J., Aruguete, D.~M., Alivisatos, A.~P. \& Geissler, P.~L.}
  2010{\natexlab{{\em a\/}}} Self-assembly of magnetic nanoparticles in
  evaporating solution. {\em J. Am. Chem. Soc.\/} {\bf 133}~(4), 838--848.

\bibitem[Ku {\em et~al.\/}(2010{\natexlab{{\em b\/}}})Ku, Aruguete, Alivisatos
  \& Geissler]{ku10}
{\sc Ku, J.-Y., Aruguete, D.~M., Alivisatos, A.~P. \& Geissler, P.~L.}
  2010{\natexlab{{\em b\/}}} Self-assembly of magnetic nanoparticles in
  evaporating solution. {\em J. Amer. Chem. Soc.\/} {\bf 133}, 838--848.

\bibitem[Love(1944)]{love}
{\sc Love, A. E.~H.} 1944 {\em A Treatise on the Mathematical Theory of
  Elasticity\/}. Dover.

\bibitem[N.Vandewalle \& S.Dorbolo(2013)]{vandewalle13}
{\sc N.Vandewalle \& S.Dorbolo} 2013 Magnetic ghosts and monopoles. {\em
  arxiv:1308.5794v1\/} .

\bibitem[Olver {\em et~al.\/}(2010)Olver, Lozier, Boisvert \& Clark]{olver10}
{\sc Olver, F. W.~J., Lozier, D.~W., Boisvert, R.~F. \& Clark, C.~W.} 2010 {\em
  {NIST} Handbook of Mathematical Functions\/}. New York: Cambridge University
  Press.

\bibitem[Perez {\em et~al.\/}(2003)Perez, Simeone, Saeki, Josephson \&
  Weissleder]{pesisa03}
{\sc Perez, J.~M., Simeone, F.~J., Saeki, Y., Josephson, L. \& Weissleder, R.}
  2003 Viral-induced self-assembly of magnetic nanoparticles allows the
  detection of viral particles in biological media. {\em Journal of the
  American Chemical Society\/} {\bf 125}~(34), 10192--10193.

\bibitem[Prokopieva {\em et~al.\/}(2009)Prokopieva, Danilov, Kantorovich \&
  Holm]{prokopieva09}
{\sc Prokopieva, T.~A., Danilov, V.~A., Kantorovich, S.~S. \& Holm, C.} 2009
  Ground state structures in ferrofluid monolayers. {\em Phys. Rev. E\/} {\bf
  80}, 031404.

\bibitem[Roper {\em et~al.\/}(2006)Roper, Dreyfus, Baudry, Fermigier, Bibette
  \& Stone]{roper06}
{\sc Roper, M., Dreyfus, R., Baudry, J., Fermigier, M., Bibette, J. \& Stone,
  H.~A.} 2006 On the dynamics of magnetically driven elastic filaments. {\em J.
  Fluid Mech.\/} {\bf 554}, 167--190.

\bibitem[Sun {\em et~al.\/}(2000)Sun, Murray, Weller, Folks \& Moser]{sumuwe00}
{\sc Sun, S., Murray, C., Weller, D., Folks, L. \& Moser, A.} 2000 Monodisperse
  {FePt} nanoparticles and ferromagnetic {FePt} nanocrystal superlattices. {\em
  Science\/} {\bf 287}~(5460), 1989--1992.

\bibitem[Var{\'o}n {\em et~al.\/}(2013)Var{\'o}n, Beleggia, Kasama, Harrison,
  Dunin-Borkowski, Puntes \& Frandsen]{vabeka13}
{\sc Var{\'o}n, M., Beleggia, M., Kasama, T., Harrison, R., Dunin-Borkowski,
  R., Puntes, V. \& Frandsen, C.} 2013 Dipolar magnetism in ordered and
  disordered low-dimensional nanoparticle assemblies. {\em Scientific
  Reports\/} {\bf 3}.

\bibitem[Wang(1986)]{wang86}
{\sc Wang, C.~Y.} 1986 A critical review of the heavy elastica. {\em Int. J.
  Mech. Sci\/} {\bf 28}, 549--559.

\bibitem[Yavuz {\em et~al.\/}(2006)Yavuz, Mayo, Yu \& Prakash]{yavuz06}
{\sc Yavuz, C.~F., Mayo, J.~T., Yu, W.~W. \& Prakash, A.} 2006 Low-field
  magnetic separation of monodisperse fe$_3$o$_4$ nanocrystals. {\em Science\/}
  {\bf 314}, 964--967.

\end{thebibliography}

\end{document}